\documentclass[12pt]{article}
\usepackage{graphicx,color,amsmath,amssymb,amsbsy}
\usepackage[utf8]{inputenc}
\usepackage[numbers,sort&compress]{natbib}
\usepackage{natbibspacing}
\usepackage[colorlinks=true, linkcolor=blue, filecolor=blue, urlcolor=blue, citecolor=blue, pdftex, plainpages=false]{hyperref}

\newcommand\pubdate{\today}
\def\CUboulder{Department of Physics, University of Colorado Boulder, Boulder, CO 80309, USA}
\def\BU{Department of Physics and Center for Computational Science, Boston University, Boston, MA 02215, USA}

\def\Title#1{\begin{center} {\Large #1 } \end{center}}
\def\Author#1{\begin{center}{ \sc #1} \end{center}}
\def\Address#1{\begin{center}{ \it #1} \end{center}}

\newcommand\pubblock{\rightline{\begin{tabular}{l} 
         \pubdate  \end{tabular}}}
\newenvironment{Abstract}{\begin{quotation}  }{\end{quotation}}
\newenvironment{Presented}{\begin{quotation} \begin{center} 
             PRESENTED AT\end{center}\bigskip 
      \begin{center}\begin{large}}{\end{large}\end{center} \end{quotation}}
\def\Acknowledgements{\bigskip  \bigskip \begin{center} \begin{large}
             \bf ACKNOWLEDGEMENTS \end{large}\end{center}}

\def\beq{\begin{equation}}
\def\eeq#1{\label{#1}\end{equation}}
\def\eeqn{\end{equation}}

\def\beqa{\begin{eqnarray}}
\def\eeqa#1{\label{#1}\end{eqnarray}}
\def\eeqan{\end{eqnarray}}

\let\bar=\overbar

\def\Dslash{\not{\hbox{\kern-4pt $D$}}}
\def\dslash{\not{\hbox{\kern-2pt $\del$}}}

\def\msb{{\bar{\ssstyle M \kern -1pt S}}}

\begin{document}
\begin{titlepage}
\pubblock

\vfill
\Title{Composite Higgs from mass-split models}
\vfill
\Author{ Oliver Witzel and Anna Hasenfratz}
\Address{\CUboulder}
\Author{ Claudio Rebbi}
\Address{\BU}
\vfill
\begin{Abstract}
Beyond Standard Model theories describing the electro-weak sector must exhibit a large separation
of scales (or ``walking'') to account for a light, 125 GeV Higgs boson and the fact that so far no
other resonances have been observed. Large separation of scales arises naturally and in a tunable
manner in mass-split models that are built on a conformal fixed point in the ultraviolet. Splitting
the fermion masses into ``light'' (massless) and ``heavy'' flavors, the system shows conformal
behavior in the ultraviolet but is chirally broken in the infrared.\\
Due to the presence of a conformal fixed point, such chirally broken systems show hyperscaling
and have a highly constrained resonance spectrum that is significantly different from the QCD
spectrum. We highlight most characteristic features presenting numerical data obtained from
dynamical simulations of an SU(3) gauge theory with four light and eight heavy flavors. In
addition, we give an outlook on ongoing work simulating an SU(3) gauge theory with four light
and six heavy flavors using a set-up well suited to explore e.g.~mass-generation of Standard Model
fermions via four-fermion interactions or partial compositeness.
\end{Abstract}
\vfill
\begin{Presented}
Thirteenth Conference on the Intersections\\ of Particle and Nuclear Physics (CIPANP2018)\\
Palm Springs, CA, USA, May 29 -- June 3, 2018
\end{Presented}
\vfill
\end{titlepage}
\def\thefootnote{\fnsymbol{footnote}}
\setcounter{footnote}{2}

\section{Introduction}
With the discovery of the Higgs boson in 2012 \cite{Aad:2012tfa,Chatrchyan:2012ufa}, all elementary particles of the Standard Model (SM) of particle physics have been confirmed experimentally. So far, however, we have not gained further insight into the nature of the Higgs boson or the origin of electro-weak symmetry breaking. Experimentally it is confirmed that the discovered Higgs boson is a scalar particle with mass 125 GeV \cite{Aad:2015zhl}  but up to the range of a few TeV, no other new resonances or particles have been observed so far. 
 
Hence we cannot say whether the discovered Higgs boson is indeed the elementary boson predicted by the SM or whether it is, e.g., of composite nature. Likewise we do not know whether new physics below the Planck-scale exists or how it looks like. Theoretically, an elementary Higgs boson would pose the challenge to UV complete the SM and issues of fine-tuning arise which can be addressed e.g.~in super-symmetric theories or theories describing the Higgs boson as composite particle.

Here we focus on exploring a class of composite Higgs models which are constructed to exhibit a large separation of scales. A large separation of scales is needed to account for the fact that a light 125 GeV Higgs boson but no other (heavier) resonances have been observed experimentally. Such other, heavier resonances are a prediction of all composite Higgs models and, given large enough energies, these resonances should be seen in collider experiments.

Schematically \cite{Ferretti:2016edi}, we want to start from the Lagrangian describing the SM without Higgs boson, ${\cal L}_{SM_0}$, and add a new sector with strong dynamics, ${\cal L}_{SD}$, as well as a term containing the interactions between SM and new strong sector, ${\cal L}_\text{int}$. With this effective ansatz, we aim to describe the physics of the SM including the Higgs boson as well as other states originating from the new strong sector and its interactions with the SM (${\cal L}_{SM} + \ldots$) 
\begin{align}
{\cal L}_{UV} \to {\cal L}_{SD} + {\cal L}_{SM_0} + {\cal L}_\text{int} \to {\cal L}_{SM} + \ldots
\end{align}
The new strong sector triggers electro-weak symmetry breaking and could give rise to a light Higgs boson. SM fermions acquire mass e.g.~through four-fermion interactions or partial compositeness. As stated above, this is an effective ansatz and a theory at some higher energy scale, ${\cal L}_{UV}$, is required to explain the origin of the new strong sector and how its fermions acquire mass. 

In the following we will solely investigate a class of models to describe ${\cal L}_{SD}$ and exhibit a large separation of scales. Since we assume this new sector features strong dynamics with nonperturbative properties similar to quantum chromodynamcis (QCD), we explore properties of ${\cal L}_{SD}$ by carrying out nonperturbative lattice field theory simulations focusing in particular at mass-split models.

\section{Mass-split models as candidates for ${\cal L}_{SD}$}
Promising candidates for ${\cal L}_{SD}$ are chirally broken in the IR but conformal in the UV \cite{Luty:2004ye,Dietrich:2006cm,Vecchi:2015fma,Ferretti:2013kya}.
\begin{figure}[!t]
  \centering
  \begin{picture}(125,20)
 \put(0,10){UV}    \put(6,11){\thicklines{\vector(1,0){110}}} \put(120,10){IR}
 \put(8,7){$\Lambda_{UV}$} \put(63,7){$\Lambda_{IR}$}
 \put(29,13){conformal} \put(25,7){fermion masses}
 \put(80,13){chirally broken} \put(80,7){Higgs dynamics}
  \end{picture}
  \caption{Sketch motivating the idea of mass-split systems. In our case the chirally broken system is given by an SU(3) gauge theory with four light (massless) flavors and we add eight (six) heavy flavors   such that our system feels the attraction of the conformal IRFP of  degenerate 12 (10) flavors.}
  \label{Fig.ScaleSep}
  \end{figure}
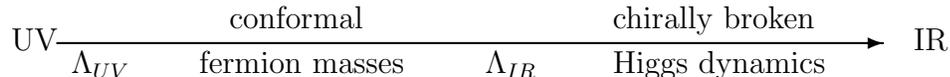
Mass-split models exhibit this feature: Starting from a conformal many flavor system in the UV (e.g.~SU(3) gauge with 10 or 12 flavors), we allow some masses to decouple in the IR and arrive at a chirally broken few flavor system. Such mass-split models are non QCD-like. While they are strongly coupled and chirally broken, dimensionless ratios of hadronic observables show conformal hyperscaling i.e.~the conformal IR fixed point (FP) governs dynamics in the UV.

In consequence, the behavior of such mass-split systems differs significantly from our experience in QCD and the role of the gauge coupling $g$ or the light and heavy flavor masses, $m_\ell$ and $m_h$ changes: 
  \begin{itemize}
  \item Physical quantities depend only on the ratio $m_\ell/m_h$
  \item The gauge coupling is irrelevant; it takes the value at the IRFP
  \item For $m_\ell \to 0$, only $m_h$ is relevant and effectively sets the scale
  \end{itemize}

  Mass-split models feature two scenarios how the Higgs boson can emerge. First, the Higgs can be a dilaton-like particle i.e.~a light pseudoscalar $0^{++}$ state which emerges from the conformal FP. For this scenario, a system with two massless flavors in the IR is ideal because that gives rise exactly to three Goldstone bosons required to give mass to the longitudinal components of the $W^\pm$ and $Z^0$ bosons of the weak interaction. The scale of the new strong interaction is set by the pseudoscalar decay constant of the new strong sector, $F_\pi$, which has to match the SM vev of about 246 GeV. Secondly, the Higgs can arise as pseudo Nambu-Goldstone boson (pNGB) \cite{Vecchi:2015fma,Ferretti:2013kya,Ferretti:2016upr,Ma:2015gra,BuarqueFranzosi:2018eaj} similar to pions in QCD. This scenario favors four massless flavors in the IR and the Higgs acquires its mass from its interaction. The vacuum alignment is non-trivial with the vacuum alignment angle $\chi$ entering the relation to set the scale $F_\pi \equiv \text{(vev SM)}/\sin(\chi) > 246$.

In the following we will present results for mass-split models with four light and eight or six heavy flavors demonstrating the general properties outlined above.  
  
\section{Results from nonperturbative simulations}
\subsection{Mass-split model with four light and eight heavy flavors}
\begin{figure}[tb]
  \centering
    \includegraphics[height=0.23\textheight]{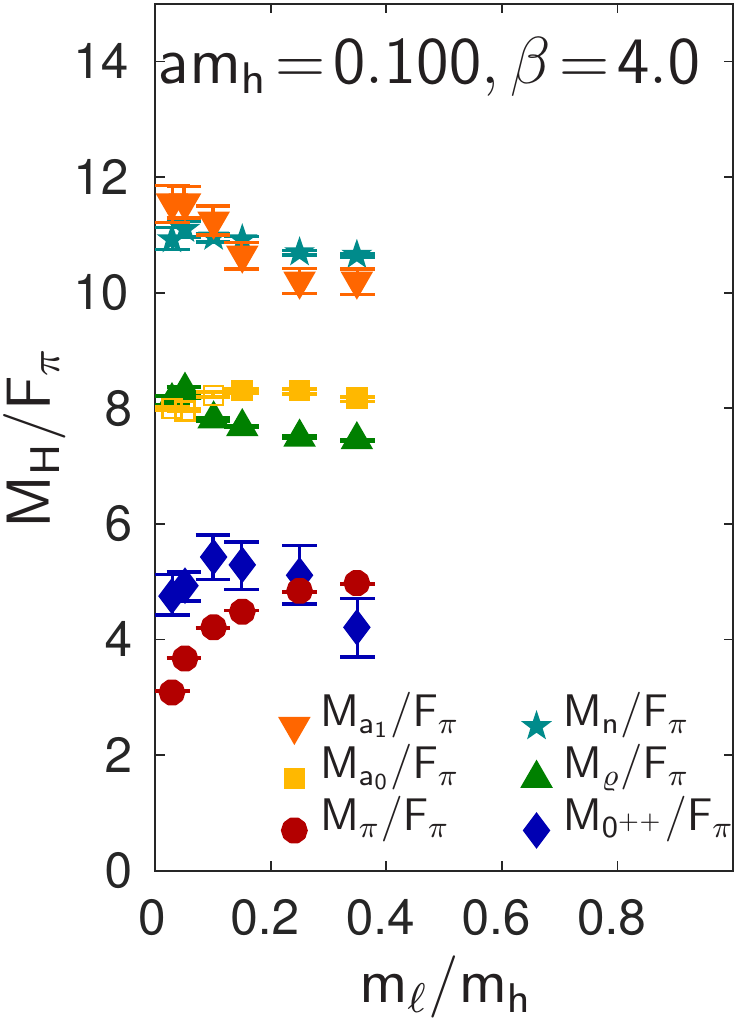}
    \includegraphics[height=0.23\textheight]{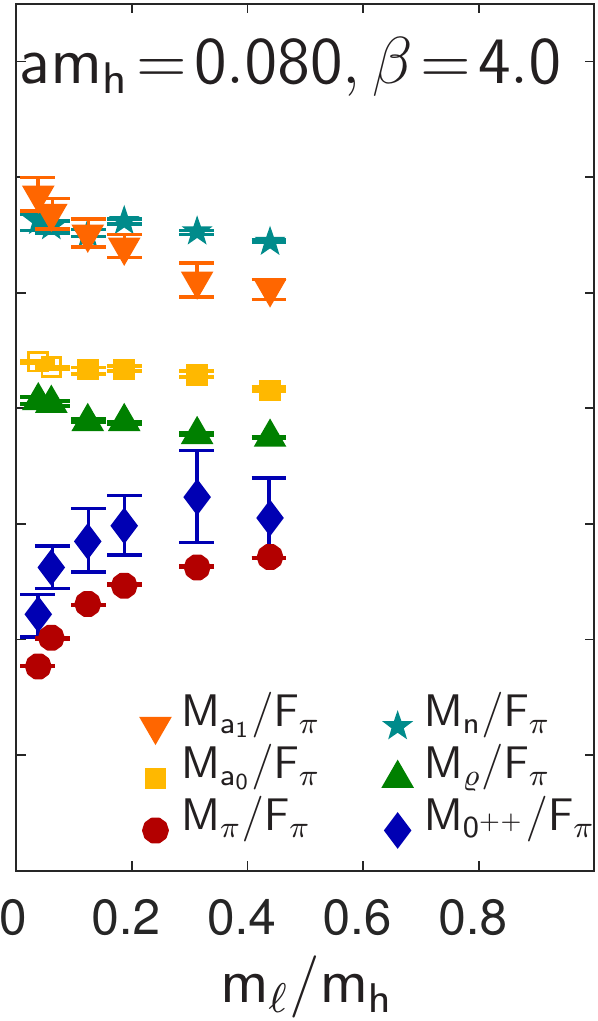}
    \includegraphics[height=0.23\textheight]{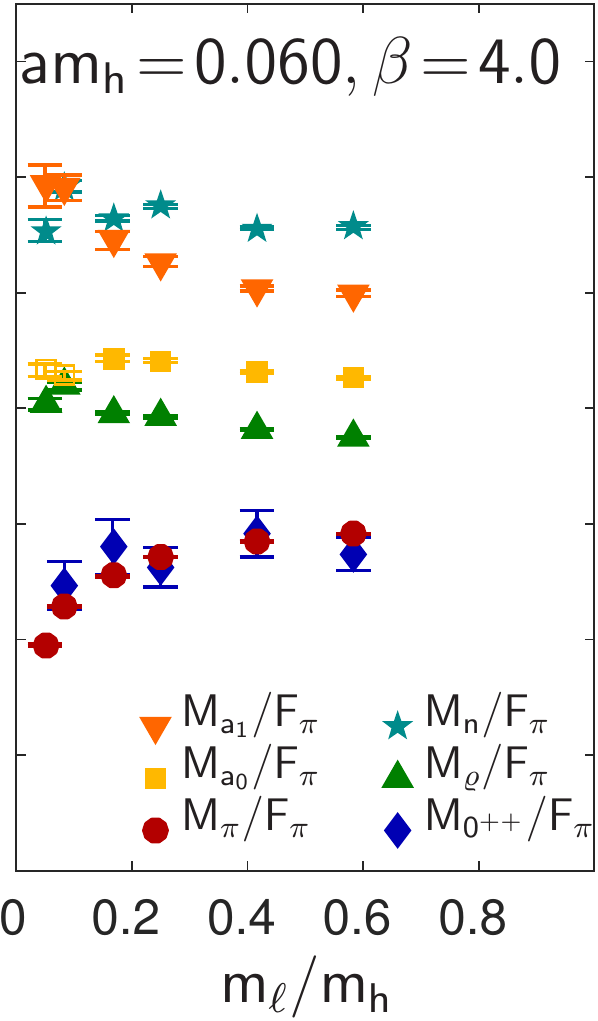}
    \includegraphics[height=0.23\textheight]{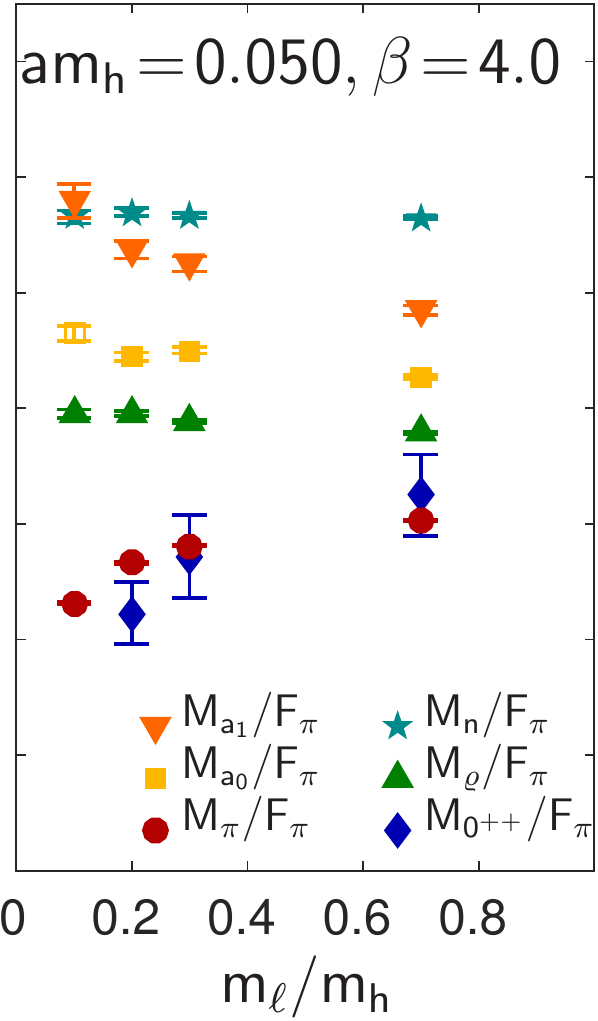}
    \includegraphics[height=0.23\textheight]{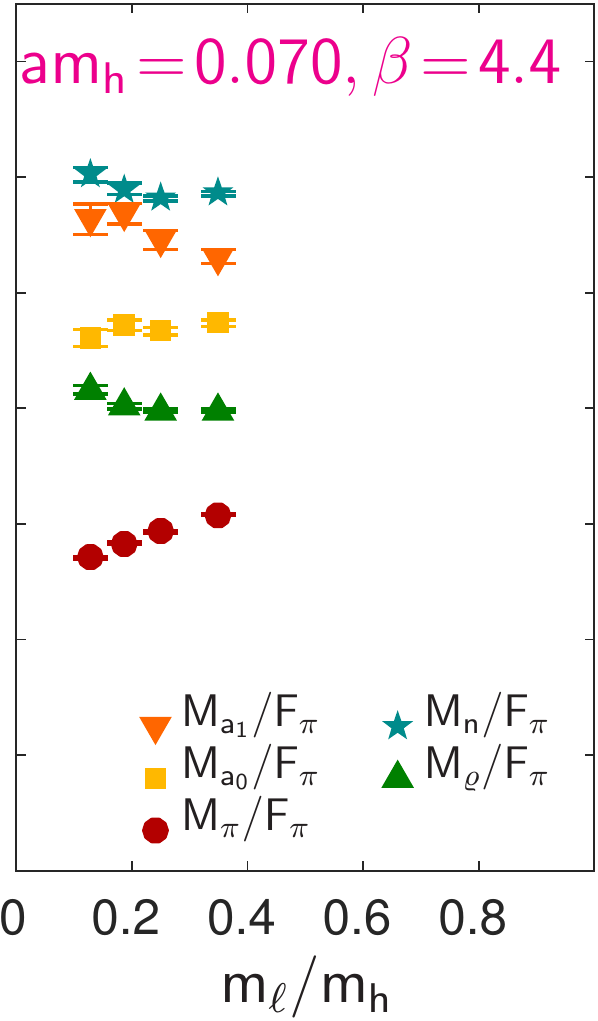}
\caption{Spectrum of the light-light bound states for five different heavy flavor masses $am_h$ at two values of the bare coupling $\beta=6/g^2$. Error bars reflect statistical uncertainties only.}
\label{fig:LightSpectrum}
\end{figure}

We simulate the system with four light and eight heavy flavors using Wilson's plaquette gauge action combined with a negative adjoint term and nHYP smeared staggered fermions \cite{Cheng:2011ic}.\footnote{Simulations are carried out using the \texttt{FUEL} code package \cite{FUEL,Osborn:2014kda}.} Ensembles are generated at two values of the bare coupling $\beta$ and five different values of the heavy flavor mass $am_h$ using up to six different values for the light flavor mass $am_\ell$ \cite{Brower:2014dfa,Brower:2014ita, Brower:2015owo,Hasenfratz:2016gut}.  In Fig.~\ref{fig:LightSpectrum} we show part of the spectrum for states made up only from light quarks. Shown are the pseudoscalar ($\pi$), iso-singlet scalar ($0^{++}$),  vector ($\rho$), scalar ($a_0$), axial ($a_1$), and nucleon ($n$) in units of the pseudoscalar decay constant $F_\pi$ vs.~the ratio of the light and heavy flavor masses, $m_\ell/m_h$. We would like to stress that only dimensionless quantities are shown avoiding any issue of scale setting.

\begin{figure}[tb]
  \centering
  \parbox{0.075\textwidth}{\includegraphics[height=0.24\textheight]{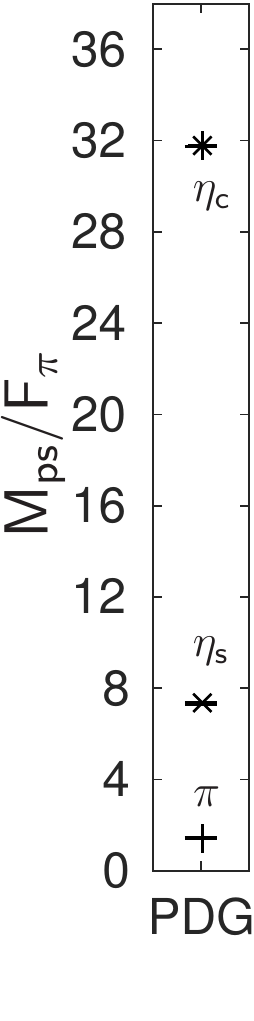}}  
  \parbox{0.19\textwidth}{\includegraphics[height=0.24\textheight]{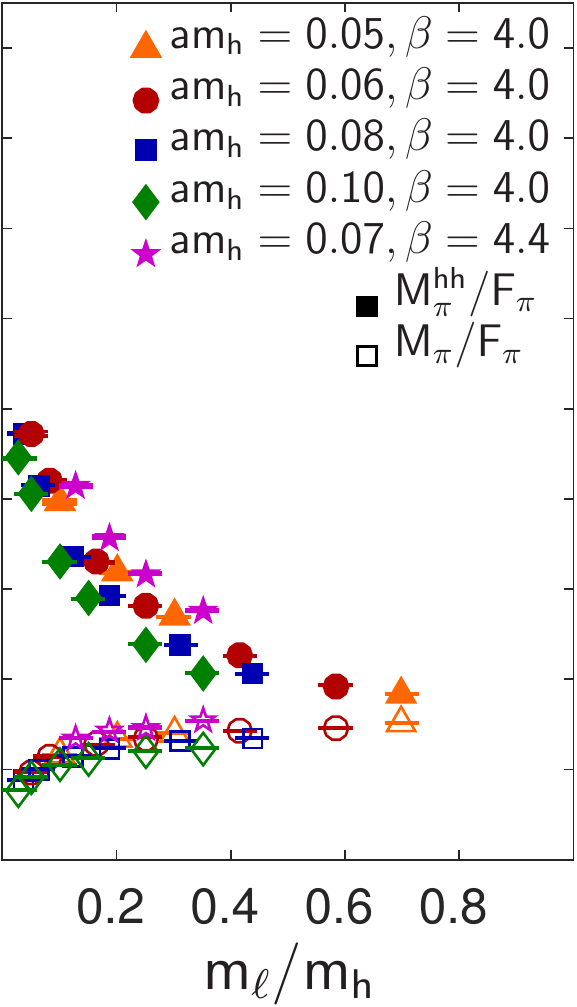}}
  \parbox{0.022\textwidth}{\includegraphics[height=0.24\textheight]{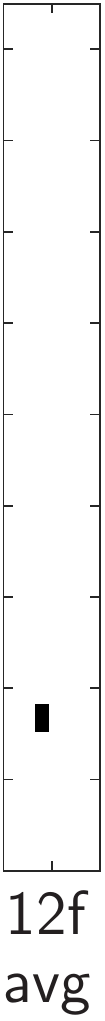}}
  \hspace{3mm}
  \parbox{0.075\textwidth}{\includegraphics[height=0.24\textheight]{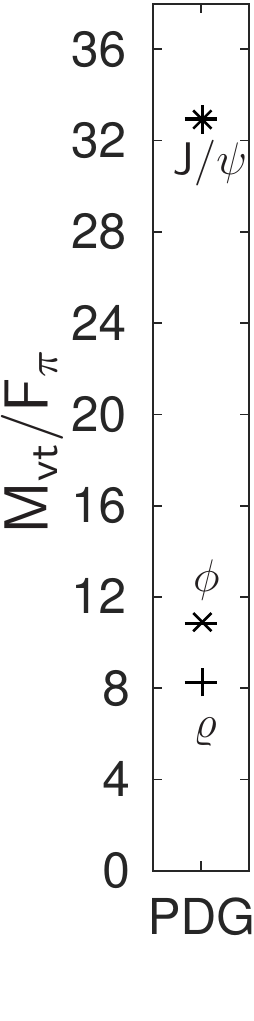}}  
  \parbox{0.19\textwidth}{\includegraphics[height=0.24\textheight]{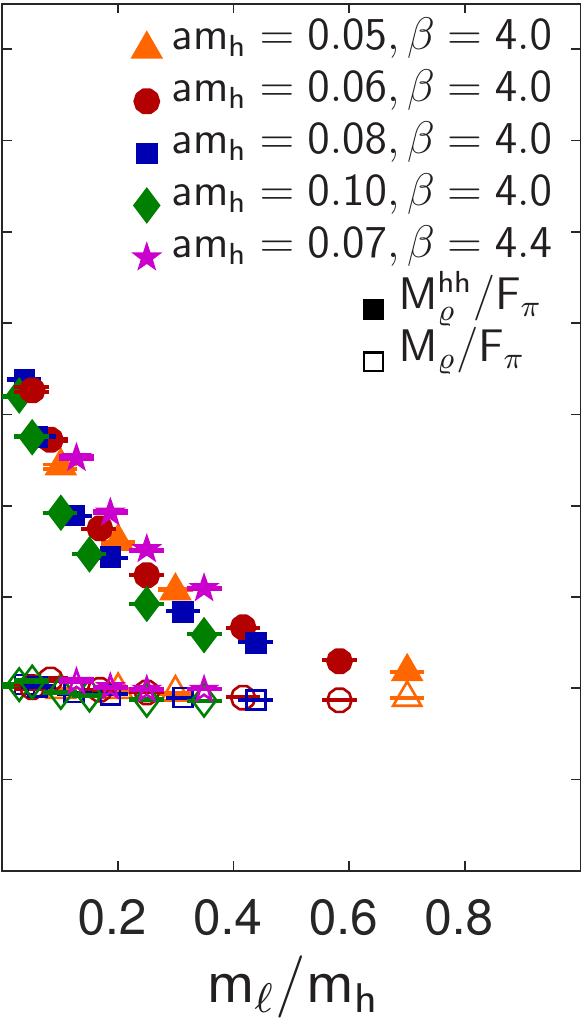}}
  \parbox{0.022\textwidth}{\includegraphics[height=0.24\textheight]{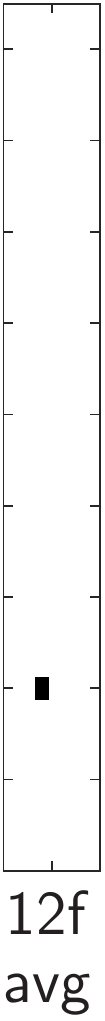}}
  \hspace{3mm}
  \parbox{0.075\textwidth}{\includegraphics[height=0.24\textheight]{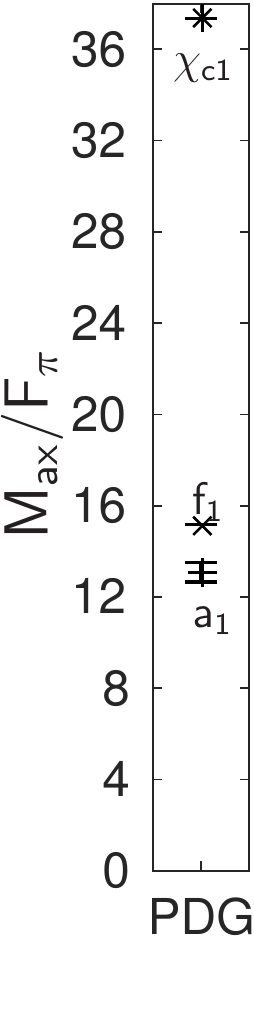}}  
  \parbox{0.19\textwidth}{\includegraphics[height=0.24\textheight]{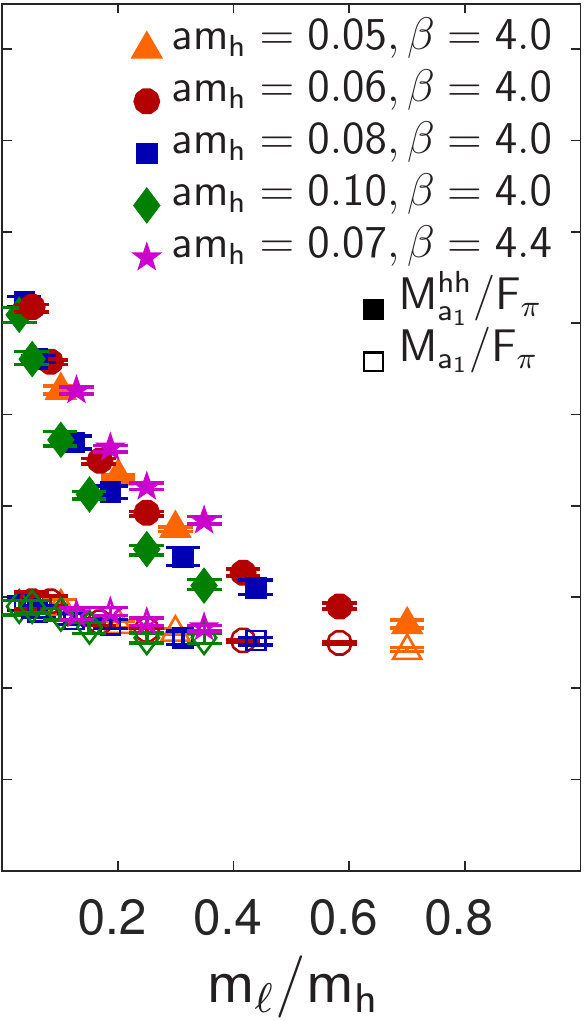}}
  \parbox{0.022\textwidth}{\includegraphics[height=0.24\textheight]{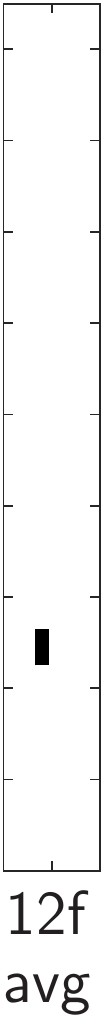}}  
  \caption{
    The three set of panels show dimensionless ratios for pseudoscalar (ps), vector (vt), and axial (ax) meson masses in units of $F_\pi$. The wide central panels show our data (with statistical errors only) as  function of $m_\ell/ m_h$. Different colors and symbols indicate the different $m_h$ and $\beta$ values, while filled (open) symbols denote states of the heavy-heavy (light-light) spectrum. The small panels to the right show averaged values for degenerate 12 flavors \cite{Aoki:2012eq,Fodor:2011tu,Cheng:2013xha,Aoki:2013zsa} and the panels on the left the corresponding PDG values \cite{Agashe:2014kda} for QCD divided by $F_\pi=94$ MeV.}
\label{fig:HyperSpectrum}
\end{figure}

Due to the contribution of (quark-line) disconnected diagrams, the calculation of the iso-singlet scalar is significantly more challenging and the obtained results have larger statistical uncertainties. Within the obtained precision, the $0^{++}$ is however degenerate with the pions and therefore seems to be quite different compared to the $\sigma$ particle (or $f_0(500)$) in the QCD spectrum. As claimed in the previous section, these dimensionless ratios in units of $F_\pi$ are independent of $m_h$ and $\beta$. This is more easily seen in Fig.~\ref{fig:HyperSpectrum} where we combine results for all ensembles in one plot and show the light-light as well as the heavy-heavy pseudoscalar, vector and axial states vs.~$m_\ell/m_h$. The different colors/symbols denote different $\beta$ or $m_h$ values; however only a unique curve as function of $m_\ell/m_h$ is traced out for each observable. For $m_\ell/m_h\to 1$ the states approach the value corresponding to degenerate 12 flavors, whereas for $m_\ell/m_h\to 0$ the fundamental differences to QCD become apparent. Quarkonia in QCD are proportional to the valence quark mass, whereas in the 4+8 model also the heavy-heavy states are independent of $m_h$.

\begin{figure}[tb]
  \centering
  \parbox{0.3\textwidth}{\includegraphics[height=0.25\textwidth]{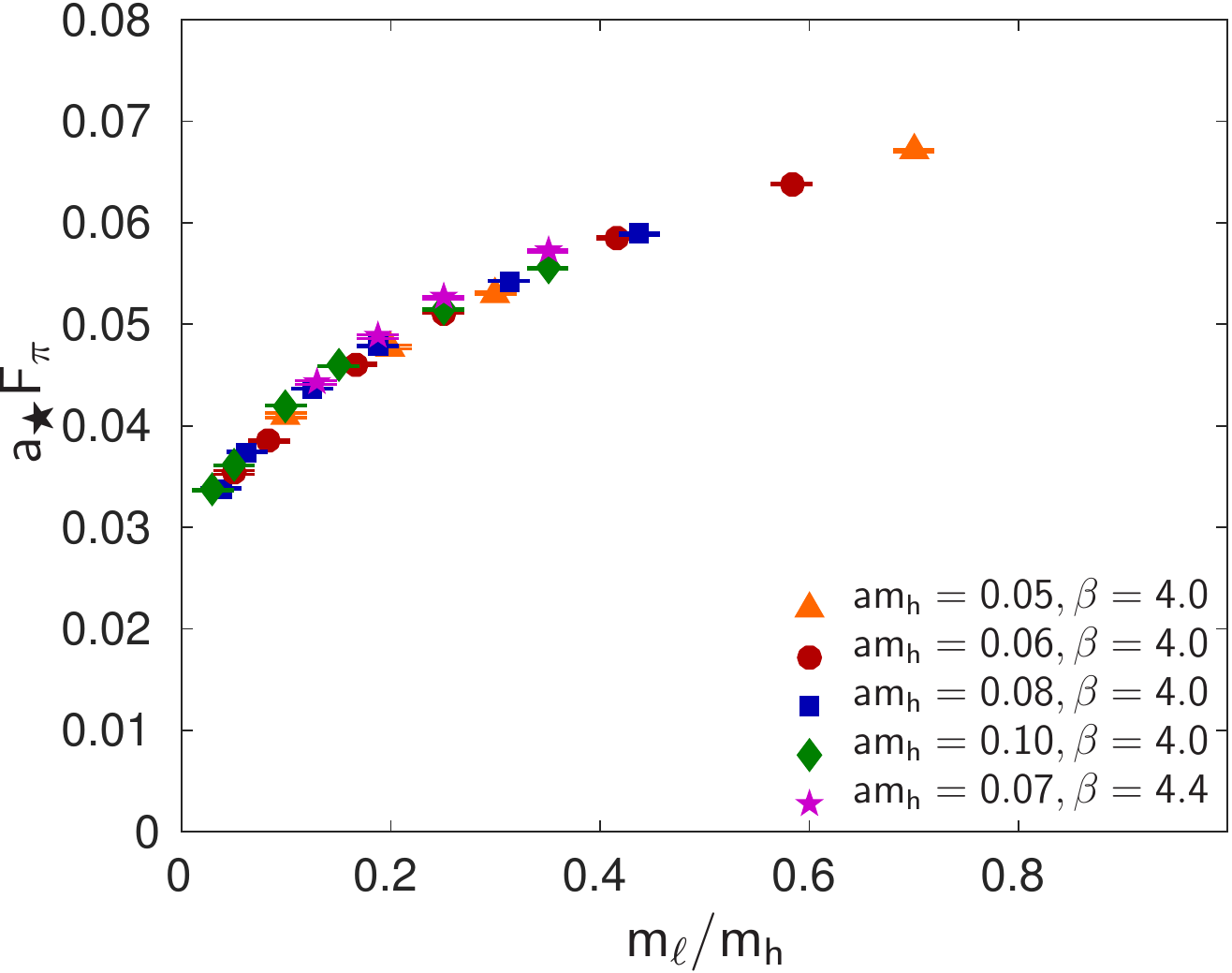}}
  \hspace{4mm}
  \parbox{0.3\textwidth}{\includegraphics[height=0.25\textwidth]{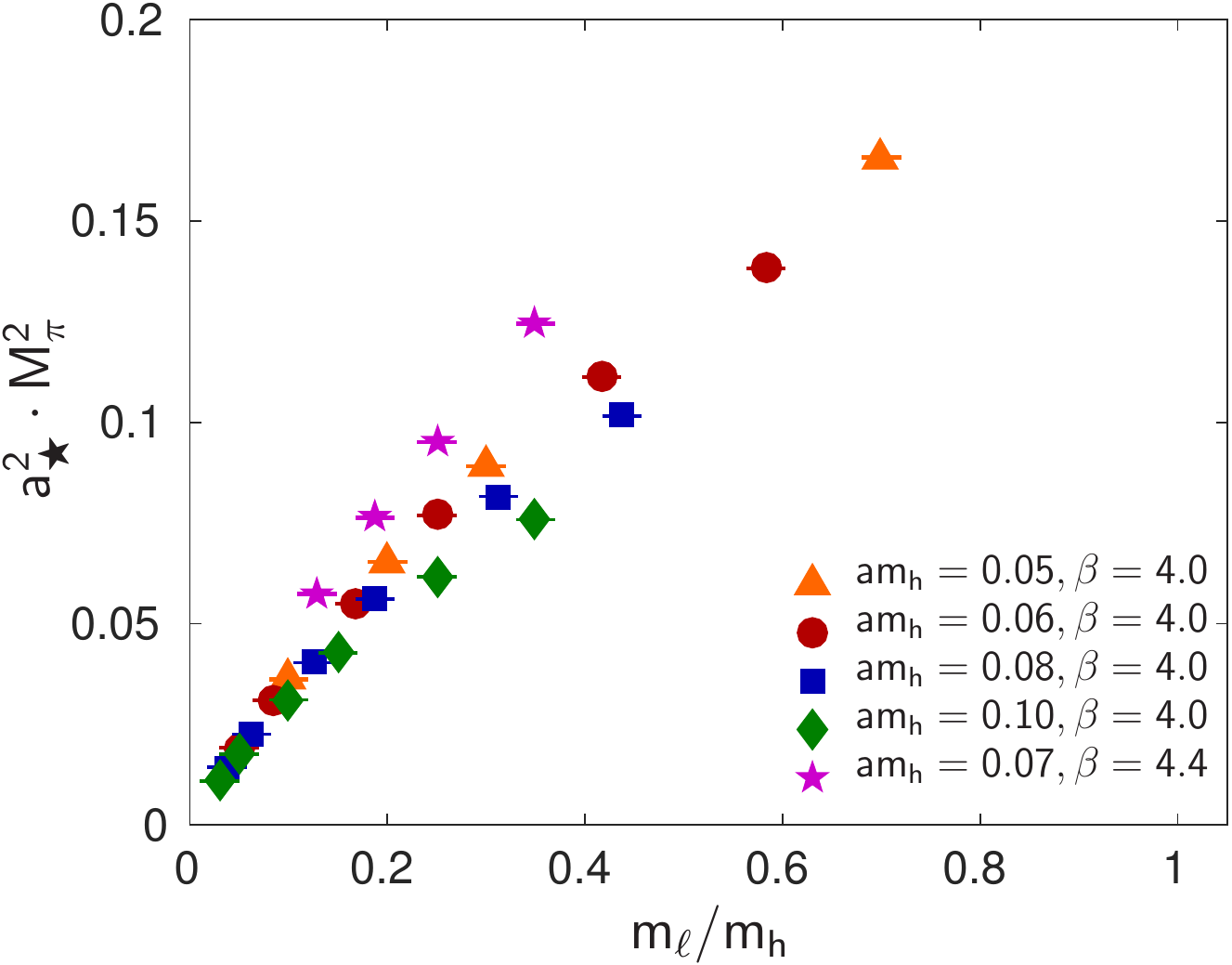}}
  \hspace{4mm}
  \parbox{0.3\textwidth}{\includegraphics[height=0.25\textwidth]{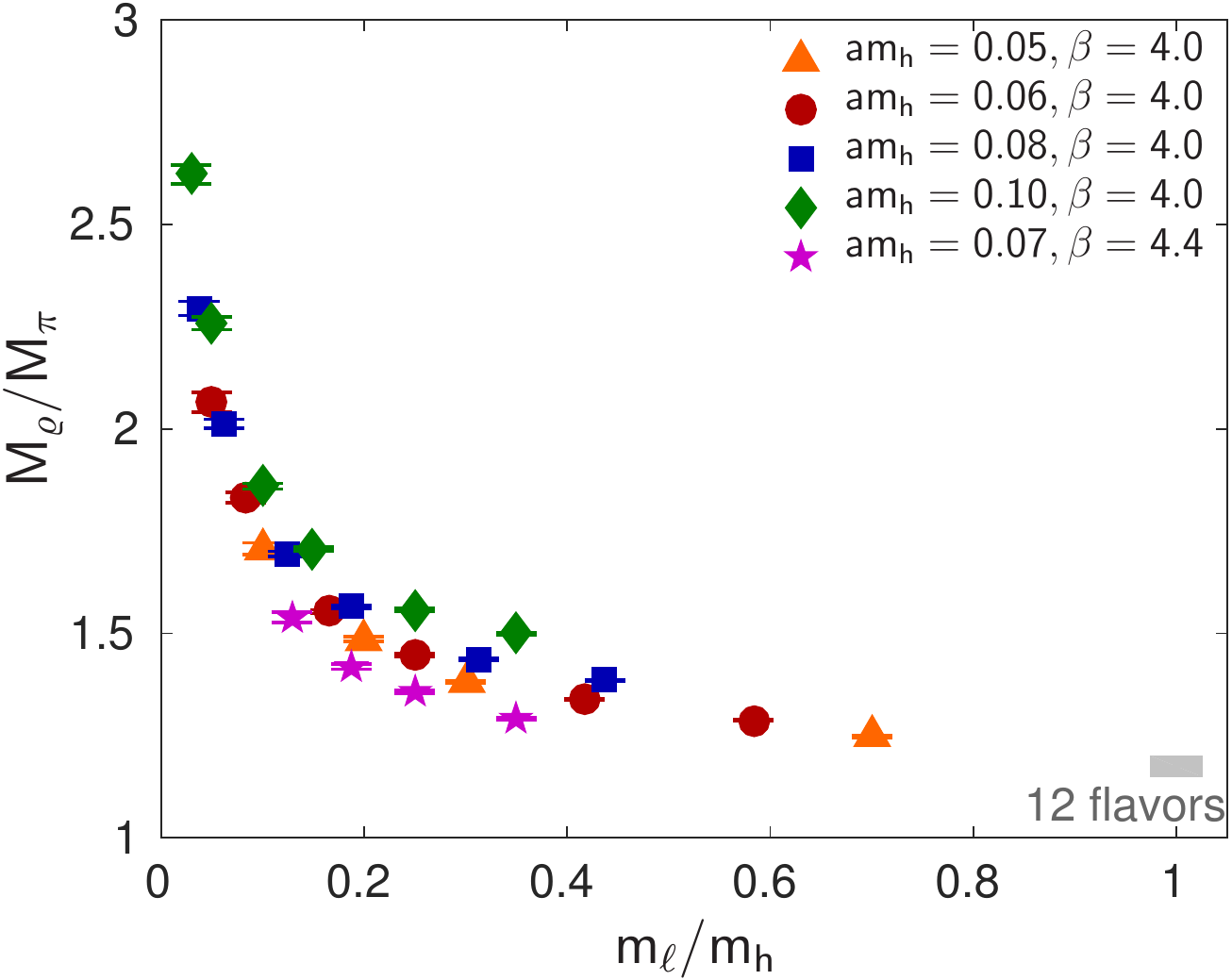}}
  \caption{The mass-split system with four light and eight heavy flavors is chirally broken: In the chiral limit ($m_\ell/m_h\to 0$), $F_\pi$ approaches a finite value and $M_\pi^2$ is proportional to $m_\ell/m_h$, while $M_\varrho/M_\pi$ diverges. Error bars reflect statistical uncertainties only.}
  \label{fig:ChirallyBroken}
\end{figure}

To demonstrate that mass-split system not only exhibit hyperscaling but are also chirally broken, we show in Fig.~\ref{fig:ChirallyBroken} the pseudoscalar decay constant $F_\pi$ (left panel) and the squared pseudoscalar meson mass $M_\pi^2$ (central panel) vs.~$m_\ell/m_h$. Here we convert results obtained on the different ensembles to the same lattice units (indicated by $a_\bigstar$) using the gradient flow scale $\sqrt{8t_0}$ \cite{Luscher:2010iy}.  As can be seen in the left panel, $F_\pi$ approaches a finite value in the chiral limit ($m_\ell/m_h\to 0$) and, for small $m_\ell/m_h$, $M_\pi^2$ decreases linearly and is proportional to $m_\ell/m_h$. For larger values of $m_\ell/m_h$ corrections to scaling and discretization effects become visible. Finally, we show in the right panel the ratio $M_\varrho/M_\pi$ which as expected diverges in the chiral limit.

\subsection{Mass-split model with four light and six heavy flavors}
As second example for a mass-split system we present preliminary results obtained as part of the Lattice Strong Dynamics Collaboration (\url{https://lsd.yale.edu/}) for an SU(3) gauge system with four light and six heavy flavors. This system is simulated using the tree-level improved Symanzik gauge action with stout-smeared M\"obius domain-wall fermions \cite{Kaneko:2013jla}.\footnote{Simulations are carried out using the \texttt{IroIro++} \cite{IroIro++} and \texttt{Grid} \cite{GRID,Boyle:2015tjk} code packages.} Domain-wall fermions feature continuum-like chiral symmetries which simplify calculations and also avoids issues of staggered fermions (e.g.~rooting, symmetry breaking \cite{Hasenfratz:2017qyr}). Reducing the number of heavy flavors from eight to six is motivated to investigate a system closer to the onset of the conformal window. If the system with ten degenerate flavors is indeed conformal, the anomalous dimension would be larger compared to a system with twelve flavors. Currently, numerical studies on the nature of $N_f=10$ are not conclusive but still ongoing \cite{Chiu:2016uui, Chiu:2017kza,Fodor:2017gtj,Hasenfratz:2017qyr}. However, observing signs of hyperscaling in mass-split simulations would be a strong indication for $N_f=10$ to be IR conformal.

\begin{figure}[tb]
  \centering
  \begin{picture}(155,70)
  \put(0,0){\includegraphics[height=0.33\textheight]{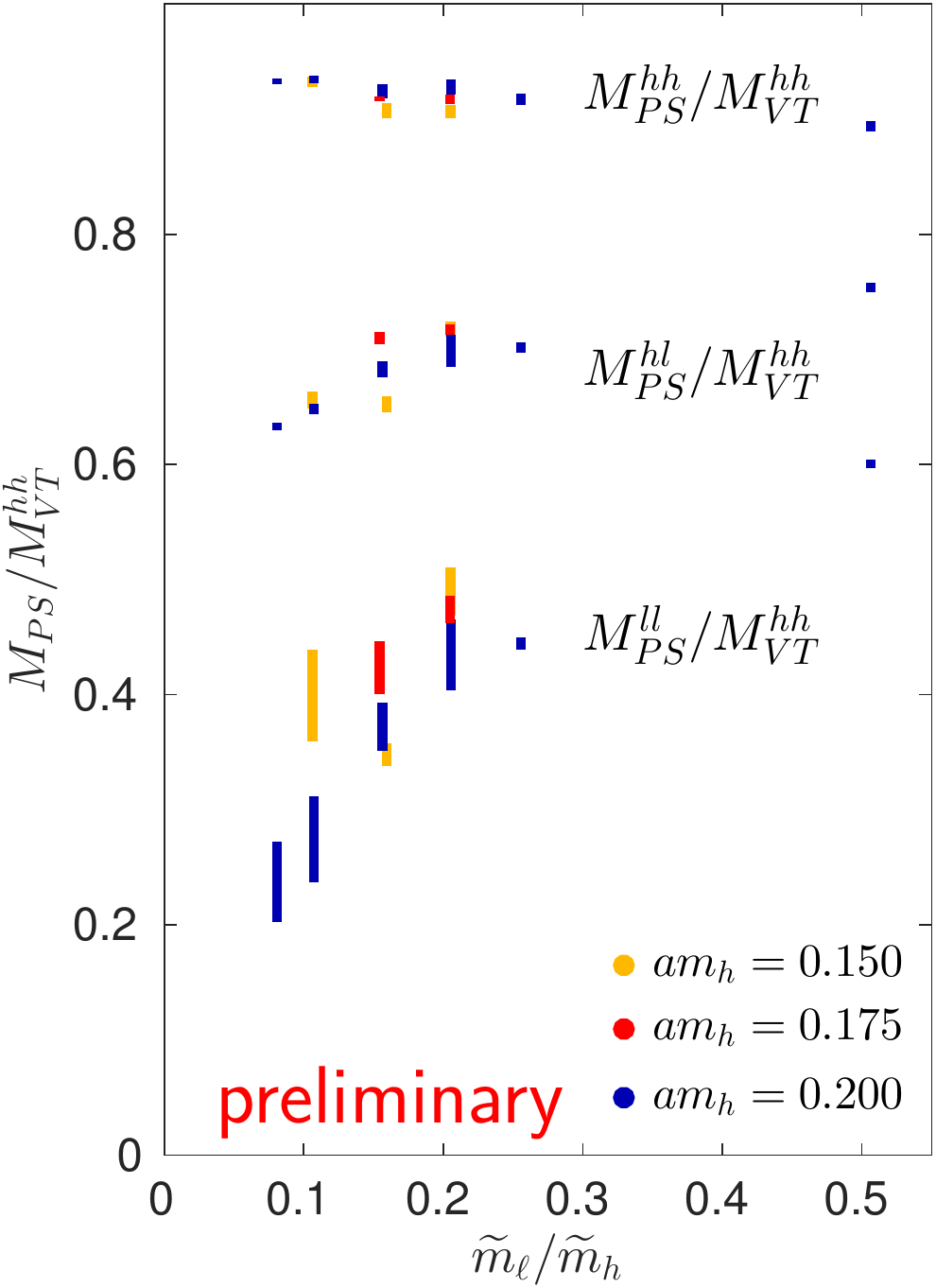}}
  \put(52,0){\includegraphics[height=0.33\textheight]{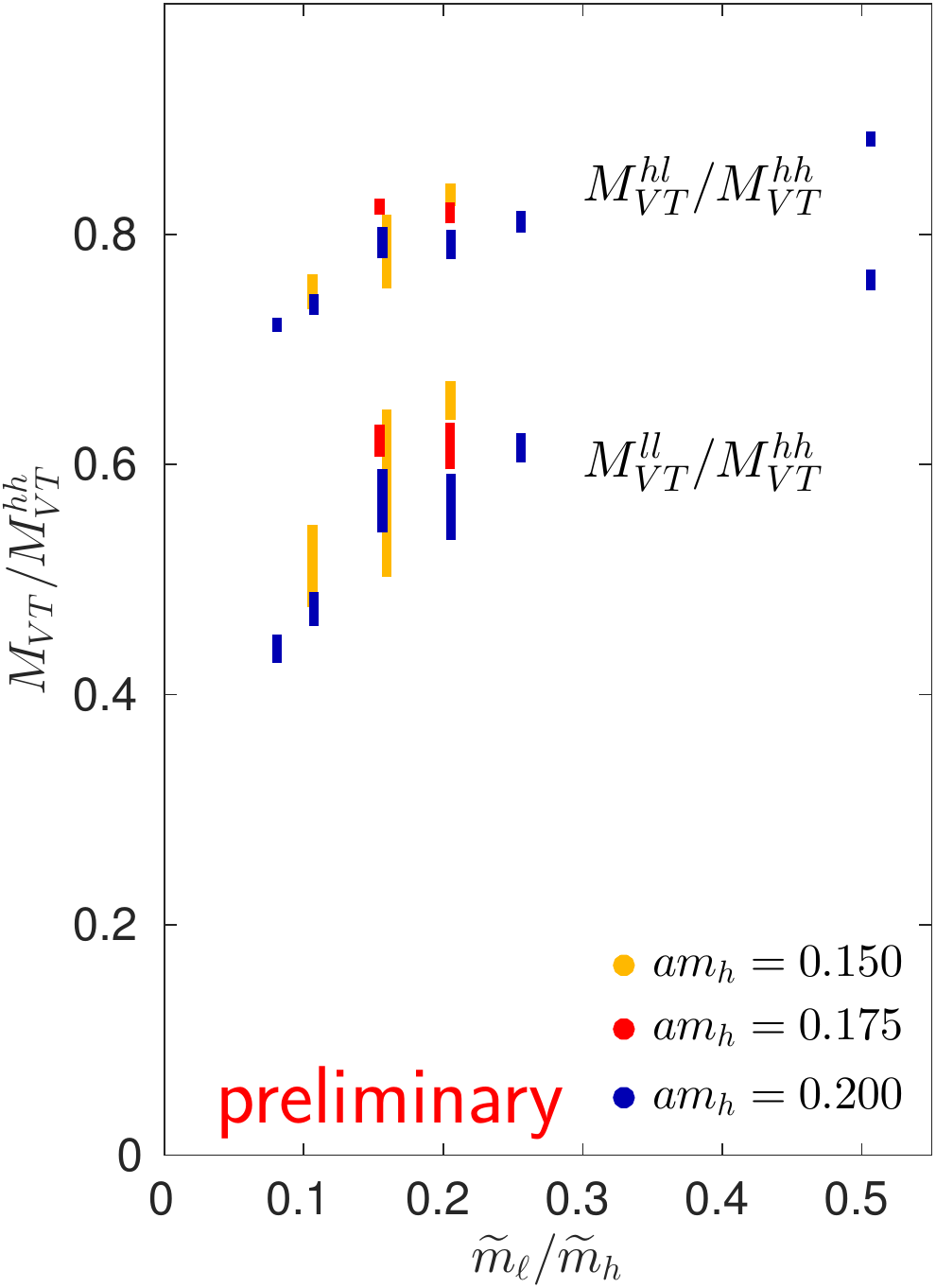}}
  \put(104,0){\includegraphics[height=0.33\textheight]{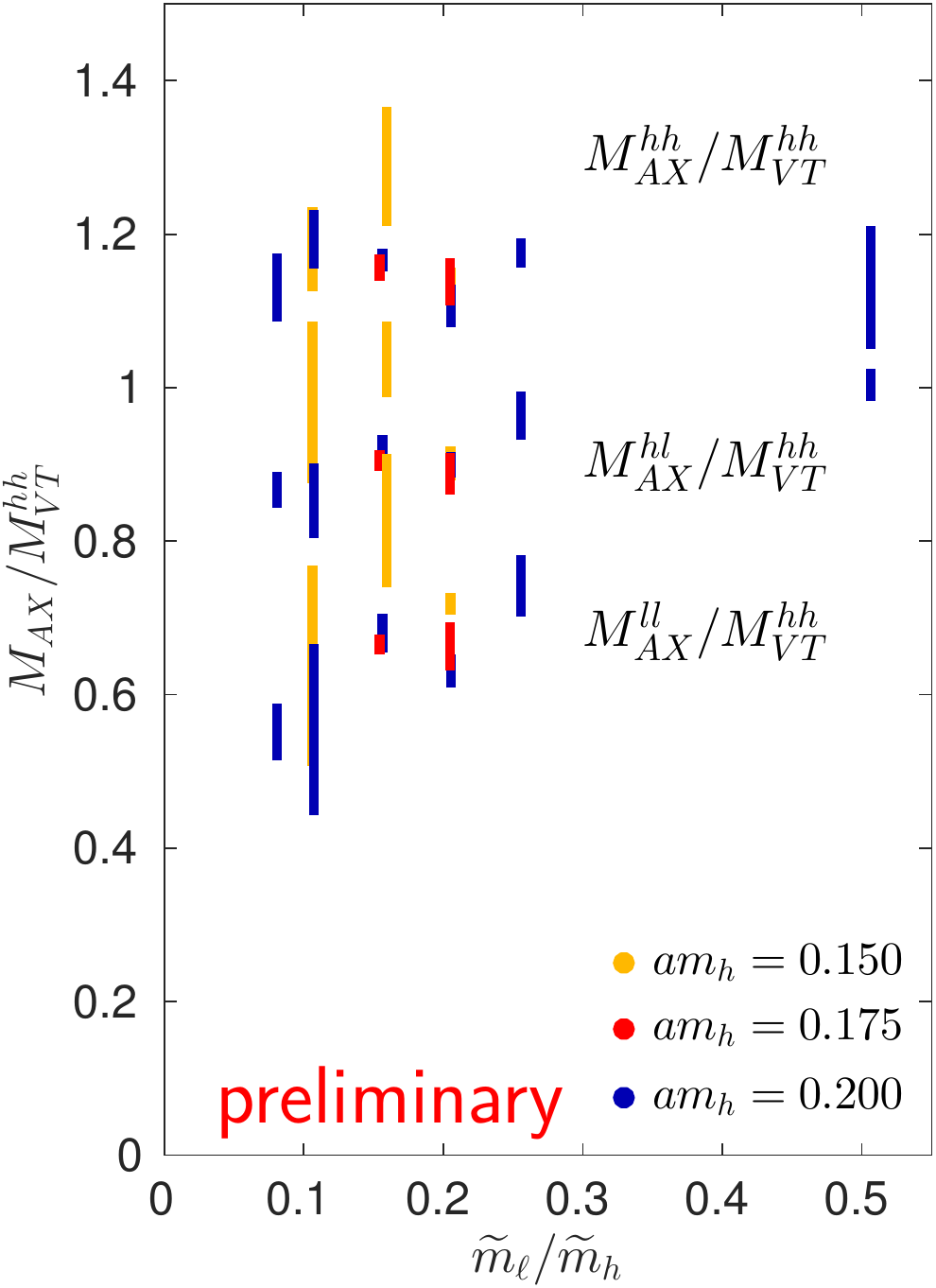}}  
  \end{picture}
  \caption{First results showing the light-light, heavy-light, and heavy-heavy pseudoscalar (left panel), vector (central panel), and axial (right panel) as ratio of the heavy-heavy vector vs.~the ratio of the flavor masses $\tilde m_\ell/\tilde m_h$. At this early stage, systematic effects may still be signifiant.}
  \label{fig:SpectrumFPS}
\end{figure}

In Fig.~\ref{fig:SpectrumFPS} we show our preliminary results for pseudoscalar, vector, and axial states. Since needed renormalization factors are still to be computed, we present the spectrum in units of the heavy-heavy vector mass vs.~$\widetilde m_\ell/\widetilde m_h$. At this early stage of the project, we have not yet obtained full control over systematic effects and also statistical uncertainties are large. We therefore only show error bands to indicate the values of the ratios but do see signs of hyperscaling emerging for the data sets featuring three different heavy quark masses. These results also indicate that we have successfully identified the parameter range of interest and are ready for more detailed investigations.

The set of ensembles is still in production and will become the basis for future investigations also targeting other quantities of pheonomenological interest like the Higgs potential, the $S$-parameter, scattering processes, or to be used for investigating partial compositeness and four-fermion interactions.

\section{Summary}
Using numerical, nonperturbative simulations we demonstrate that mass-split models in the basin of attraction of an IRFP exhibit unique properties which are rather different from QCD-like systems. By construction, these models exhibit a large scale separation and due to inheriting hyperscaling they are also highly predictive. Hence they are viable candidates for a new strong sector introduced with the aim to explain the origin of electro-weak symmetry breaking and the Higgs boson as composite particle. 

Numerically we find that the iso-singlet scalar ($0^{++}$) is light and degenerate with the pseudoscalar (pion). In order to better understand mass-split models, to better explore their properties, and calculate phenomenologically testable observables, we have together with the Lattice Strong Dynamics collaboration started to explore a system with four light and six heavy flavors using conceptually clean domain-wall fermions which will simplify further studies in the future.

\Acknowledgements
The authors thank their colleagues in the LSD Collaboration for fruitful and inspiring discussions. We are grateful to Peter Boyle, Guido Cossu, James Osborn, Antonin Portelli, and Azusa Yamaguchi for developing software providing the basis of this work.
Computations for this work were carried out in part on facilities of the USQCD Collaboration, which are funded by the Office of Science of the U.S.~Department of Energy, on computers at the MGHPCC, in part funded by the National Science Foundation (award OCI-1229059), and on computers allocated under the NSF Xsede program to the project TG-PHY120002. We thank Boston University, Fermilab, Jefferson Lab, the NSF and the U.S.~DOE for providing the facilities essential for the completion of this work.  Further we thank the Lawrence Livermore National Laboratory (LLNL) Multiprogrammatic and Institutional Computing program for Grand Challenge allocations and time on the LLNL BlueGene/Q supercomputer. A.H.~and O.W.~acknowledge support by the Office of Science of the Department of Energy grant DE-SC0010005, C.R.~acknowleges support under  DOE grant DE-SC0015845.

\bibliography{../General/BSM}

\begin{thebibliography}{32}%
\makeatletter
\providecommand \@ifxundefined [1]{%
 \@ifx{#1\undefined}
}%
\providecommand \@ifnum [1]{%
 \ifnum #1\expandafter \@firstoftwo
 \else \expandafter \@secondoftwo
 \fi
}%
\providecommand \@ifx [1]{%
 \ifx #1\expandafter \@firstoftwo
 \else \expandafter \@secondoftwo
 \fi
}%
\providecommand \natexlab [1]{#1}%
\providecommand \enquote  [1]{``#1''}%
\providecommand \bibnamefont  [1]{#1}%
\providecommand \bibfnamefont [1]{#1}%
\providecommand \citenamefont [1]{#1}%
\providecommand \href@noop [0]{\@secondoftwo}%
\providecommand \href [0]{\begingroup \@sanitize@url \@href}%
\providecommand \@href[1]{\@@startlink{#1}\@@href}%
\providecommand \@@href[1]{\endgroup#1\@@endlink}%
\providecommand \@sanitize@url [0]{\catcode `\\12\catcode `\$12\catcode
  `\&12\catcode `\#12\catcode `\^12\catcode `\_12\catcode `\%12\relax}%
\providecommand \@@startlink[1]{}%
\providecommand \@@endlink[0]{}%
\providecommand \url  [0]{\begingroup\@sanitize@url \@url }%
\providecommand \@url [1]{\endgroup\@href {#1}{\urlprefix }}%
\providecommand \urlprefix  [0]{URL }%
\providecommand \Eprint [0]{\href }%
\providecommand \doibase [0]{http://dx.doi.org/}%
\providecommand \selectlanguage [0]{\@gobble}%
\providecommand \bibinfo  [0]{\@secondoftwo}%
\providecommand \bibfield  [0]{\@secondoftwo}%
\providecommand \translation [1]{[#1]}%
\providecommand \BibitemOpen [0]{}%
\providecommand \bibitemStop [0]{}%
\providecommand \bibitemNoStop [0]{.\EOS\space}%
\providecommand \EOS [0]{\spacefactor3000\relax}%
\providecommand \BibitemShut  [1]{\csname bibitem#1\endcsname}%
\let\auto@bib@innerbib\@empty
\bibitem [{\citenamefont {Aad}\ \emph {et~al.}(2012)\citenamefont {Aad} \emph
  {et~al.}}]{Aad:2012tfa}%
  \BibitemOpen
  \bibfield  {author} {\bibinfo {author} {\bibfnamefont {G.}~\bibnamefont
  {Aad}} \emph {et~al.} (\bibinfo {collaboration} {ATLAS}),\ }\href {\doibase
  10.1016/j.physletb.2012.08.020} {\bibfield  {journal} {\bibinfo  {journal}
  {Phys.Lett.}\ }\textbf {\bibinfo {volume} {B716}},\ \bibinfo {pages} {1}
  (\bibinfo {year} {2012})},\ \Eprint {http://arxiv.org/abs/1207.7214}
  {arXiv:1207.7214 [hep-ex]} \BibitemShut {NoStop}%
\bibitem [{\citenamefont {Chatrchyan}\ \emph {et~al.}(2012)\citenamefont
  {Chatrchyan} \emph {et~al.}}]{Chatrchyan:2012ufa}%
  \BibitemOpen
  \bibfield  {author} {\bibinfo {author} {\bibfnamefont {S.}~\bibnamefont
  {Chatrchyan}} \emph {et~al.} (\bibinfo {collaboration} {CMS Collaboration}),\
  }\href {\doibase 10.1016/j.physletb.2012.08.021} {\bibfield  {journal}
  {\bibinfo  {journal} {Phys.Lett.}\ }\textbf {\bibinfo {volume} {B716}},\
  \bibinfo {pages} {30} (\bibinfo {year} {2012})},\ \Eprint
  {http://arxiv.org/abs/1207.7235} {arXiv:1207.7235 [hep-ex]} \BibitemShut
  {NoStop}%
\bibitem [{\citenamefont {Aad}\ \emph {et~al.}(2015)\citenamefont {Aad} \emph
  {et~al.}}]{Aad:2015zhl}%
  \BibitemOpen
  \bibfield  {author} {\bibinfo {author} {\bibfnamefont {G.}~\bibnamefont
  {Aad}} \emph {et~al.} (\bibinfo {collaboration} {ATLAS, CMS}),\ }\href
  {\doibase 10.1103/PhysRevLett.114.191803} {\bibfield  {journal} {\bibinfo
  {journal} {Phys. Rev. Lett.}\ }\textbf {\bibinfo {volume} {114}},\ \bibinfo
  {pages} {191803} (\bibinfo {year} {2015})},\ \Eprint
  {http://arxiv.org/abs/1503.07589} {arXiv:1503.07589 [hep-ex]} \BibitemShut
  {NoStop}%
\bibitem [{\citenamefont {Ferretti}()}]{Ferretti:2016edi}%
  \BibitemOpen
  \bibfield  {author} {\bibinfo {author} {\bibfnamefont {G.}~\bibnamefont
  {Ferretti}},\ }\href
  {{https://indico.ph.ed.ac.uk/event/18/contribution/10/material/slides/0.pdf}}
  {}\bibinfo {note} {{Talk at the workshop ``Holography, conformal field
  theories, and lattice'' in Edinburgh, Scotland}}\BibitemShut {NoStop}%
\bibitem [{\citenamefont {Luty}\ and\ \citenamefont
  {Okui}(2006)}]{Luty:2004ye}%
  \BibitemOpen
  \bibfield  {author} {\bibinfo {author} {\bibfnamefont {M.~A.}\ \bibnamefont
  {Luty}}\ and\ \bibinfo {author} {\bibfnamefont {T.}~\bibnamefont {Okui}},\
  }\href {\doibase 10.1088/1126-6708/2006/09/070} {\bibfield  {journal}
  {\bibinfo  {journal} {JHEP}\ }\textbf {\bibinfo {volume} {09}},\ \bibinfo
  {pages} {070} (\bibinfo {year} {2006})},\ \Eprint
  {http://arxiv.org/abs/hep-ph/0409274} {arXiv:hep-ph/0409274 [hep-ph]}
  \BibitemShut {NoStop}%
\bibitem [{\citenamefont {Dietrich}\ and\ \citenamefont
  {Sannino}(2007)}]{Dietrich:2006cm}%
  \BibitemOpen
  \bibfield  {author} {\bibinfo {author} {\bibfnamefont {D.~D.}\ \bibnamefont
  {Dietrich}}\ and\ \bibinfo {author} {\bibfnamefont {F.}~\bibnamefont
  {Sannino}},\ }\href {\doibase 10.1103/PhysRevD.75.085018} {\bibfield
  {journal} {\bibinfo  {journal} {Phys. Rev.}\ }\textbf {\bibinfo {volume}
  {D75}},\ \bibinfo {pages} {085018} (\bibinfo {year} {2007})},\ \Eprint
  {http://arxiv.org/abs/hep-ph/0611341} {arXiv:hep-ph/0611341 [hep-ph]}
  \BibitemShut {NoStop}%
\bibitem [{\citenamefont {Vecchi}(2015)}]{Vecchi:2015fma}%
  \BibitemOpen
  \bibfield  {author} {\bibinfo {author} {\bibfnamefont {L.}~\bibnamefont
  {Vecchi}},\ }\href@noop {} {\  (\bibinfo {year} {2015})},\ \Eprint
  {http://arxiv.org/abs/1506.00623} {arXiv:1506.00623 [hep-ph]} \BibitemShut
  {NoStop}%
\bibitem [{\citenamefont {Ferretti}\ and\ \citenamefont
  {Karateev}(2014)}]{Ferretti:2013kya}%
  \BibitemOpen
  \bibfield  {author} {\bibinfo {author} {\bibfnamefont {G.}~\bibnamefont
  {Ferretti}}\ and\ \bibinfo {author} {\bibfnamefont {D.}~\bibnamefont
  {Karateev}},\ }\href {\doibase 10.1007/JHEP03(2014)077} {\bibfield  {journal}
  {\bibinfo  {journal} {JHEP}\ }\textbf {\bibinfo {volume} {03}},\ \bibinfo
  {pages} {077} (\bibinfo {year} {2014})},\ \Eprint
  {http://arxiv.org/abs/1312.5330} {arXiv:1312.5330 [hep-ph]} \BibitemShut
  {NoStop}%
\bibitem [{\citenamefont {Ferretti}(2016)}]{Ferretti:2016upr}%
  \BibitemOpen
  \bibfield  {author} {\bibinfo {author} {\bibfnamefont {G.}~\bibnamefont
  {Ferretti}},\ }\href {\doibase 10.1007/JHEP06(2016)107} {\bibfield  {journal}
  {\bibinfo  {journal} {JHEP}\ }\textbf {\bibinfo {volume} {06}},\ \bibinfo
  {pages} {107} (\bibinfo {year} {2016})},\ \Eprint
  {http://arxiv.org/abs/1604.06467} {arXiv:1604.06467 [hep-ph]} \BibitemShut
  {NoStop}%
\bibitem [{\citenamefont {Ma}\ and\ \citenamefont
  {Cacciapaglia}(2016)}]{Ma:2015gra}%
  \BibitemOpen
  \bibfield  {author} {\bibinfo {author} {\bibfnamefont {T.}~\bibnamefont
  {Ma}}\ and\ \bibinfo {author} {\bibfnamefont {G.}~\bibnamefont
  {Cacciapaglia}},\ }\href {\doibase 10.1007/JHEP03(2016)211} {\bibfield
  {journal} {\bibinfo  {journal} {JHEP}\ }\textbf {\bibinfo {volume} {03}},\
  \bibinfo {pages} {211} (\bibinfo {year} {2016})},\ \Eprint
  {http://arxiv.org/abs/1508.07014} {arXiv:1508.07014 [hep-ph]} \BibitemShut
  {NoStop}%
\bibitem [{\citenamefont {Buarque~Franzosi}\ \emph {et~al.}(2018)\citenamefont
  {Buarque~Franzosi}, \citenamefont {Cacciapaglia},\ and\ \citenamefont
  {Deandrea}}]{BuarqueFranzosi:2018eaj}%
  \BibitemOpen
  \bibfield  {author} {\bibinfo {author} {\bibfnamefont {D.}~\bibnamefont
  {Buarque~Franzosi}}, \bibinfo {author} {\bibfnamefont {G.}~\bibnamefont
  {Cacciapaglia}}, \ and\ \bibinfo {author} {\bibfnamefont {A.}~\bibnamefont
  {Deandrea}},\ }\href@noop {} {\  (\bibinfo {year} {2018})},\ \Eprint
  {http://arxiv.org/abs/1809.09146} {arXiv:1809.09146 [hep-ph]} \BibitemShut
  {NoStop}%
\bibitem [{\citenamefont {Cheng}\ \emph {et~al.}(2012)\citenamefont {Cheng},
  \citenamefont {Hasenfratz},\ and\ \citenamefont {Schaich}}]{Cheng:2011ic}%
  \BibitemOpen
  \bibfield  {author} {\bibinfo {author} {\bibfnamefont {A.}~\bibnamefont
  {Cheng}}, \bibinfo {author} {\bibfnamefont {A.}~\bibnamefont {Hasenfratz}}, \
  and\ \bibinfo {author} {\bibfnamefont {D.}~\bibnamefont {Schaich}},\ }\href
  {\doibase 10.1103/PhysRevD.85.094509} {\bibfield  {journal} {\bibinfo
  {journal} {Phys.Rev.}\ }\textbf {\bibinfo {volume} {D85}},\ \bibinfo {pages}
  {094509} (\bibinfo {year} {2012})},\ \Eprint {http://arxiv.org/abs/1111.2317}
  {arXiv:1111.2317 [hep-lat]} \BibitemShut {NoStop}%
\bibitem [{\citenamefont {Osborn}\ \emph {et~al.}()\citenamefont {Osborn} \emph
  {et~al.}}]{FUEL}%
  \BibitemOpen
  \bibfield  {author} {\bibinfo {author} {\bibfnamefont {J.}~\bibnamefont
  {Osborn}} \emph {et~al.},\ }\href {http://usqcd-software.github.io/FUEL.html}
  {\enquote {\bibinfo {title} {{Framework for unified evolution of lattices
  (FUEL)}},}\ }\BibitemShut {NoStop}%
\bibitem [{\citenamefont {Osborn}(2014)}]{Osborn:2014kda}%
  \BibitemOpen
  \bibfield  {author} {\bibinfo {author} {\bibfnamefont {J.}~\bibnamefont
  {Osborn}},\ }\href@noop {} {\bibfield  {journal} {\bibinfo  {journal} {PoS}\
  }\textbf {\bibinfo {volume} {LATTICE2014}},\ \bibinfo {pages} {028} (\bibinfo
  {year} {2014})}\BibitemShut {NoStop}%
\bibitem [{\citenamefont {Brower}\ \emph {et~al.}(2015)\citenamefont {Brower},
  \citenamefont {Hasenfratz}, \citenamefont {Rebbi}, \citenamefont {Weinberg},\
  and\ \citenamefont {Witzel}}]{Brower:2014dfa}%
  \BibitemOpen
  \bibfield  {author} {\bibinfo {author} {\bibfnamefont {R.~C.}\ \bibnamefont
  {Brower}}, \bibinfo {author} {\bibfnamefont {A.}~\bibnamefont {Hasenfratz}},
  \bibinfo {author} {\bibfnamefont {C.}~\bibnamefont {Rebbi}}, \bibinfo
  {author} {\bibfnamefont {E.}~\bibnamefont {Weinberg}}, \ and\ \bibinfo
  {author} {\bibfnamefont {O.}~\bibnamefont {Witzel}},\ }\href {\doibase
  10.1134/S1063776115030176} {\bibfield  {journal} {\bibinfo  {journal} {J.
  Exp. Theor. Phys.}\ }\textbf {\bibinfo {volume} {120}},\ \bibinfo {pages}
  {423} (\bibinfo {year} {2015})},\ \Eprint {http://arxiv.org/abs/1410.4091}
  {arXiv:1410.4091 [hep-lat]} \BibitemShut {NoStop}%
\bibitem [{\citenamefont {Brower}\ \emph {et~al.}(2014)\citenamefont {Brower},
  \citenamefont {Hasenfratz}, \citenamefont {Rebbi}, \citenamefont {Weinberg},\
  and\ \citenamefont {Witzel}}]{Brower:2014ita}%
  \BibitemOpen
  \bibfield  {author} {\bibinfo {author} {\bibfnamefont {R.~C.}\ \bibnamefont
  {Brower}}, \bibinfo {author} {\bibfnamefont {A.}~\bibnamefont {Hasenfratz}},
  \bibinfo {author} {\bibfnamefont {C.}~\bibnamefont {Rebbi}}, \bibinfo
  {author} {\bibfnamefont {E.}~\bibnamefont {Weinberg}}, \ and\ \bibinfo
  {author} {\bibfnamefont {O.}~\bibnamefont {Witzel}},\ }\href@noop {}
  {\bibfield  {journal} {\bibinfo  {journal} {PoS}\ }\textbf {\bibinfo {volume}
  {LATTICE2014}},\ \bibinfo {pages} {254} (\bibinfo {year} {2014})},\ \Eprint
  {http://arxiv.org/abs/1411.3243} {arXiv:1411.3243 [hep-lat]} \BibitemShut
  {NoStop}%
\bibitem [{\citenamefont {Brower}\ \emph {et~al.}(2016)\citenamefont {Brower},
  \citenamefont {Hasenfratz}, \citenamefont {Rebbi}, \citenamefont {Weinberg},\
  and\ \citenamefont {Witzel}}]{Brower:2015owo}%
  \BibitemOpen
  \bibfield  {author} {\bibinfo {author} {\bibfnamefont {R.~C.}\ \bibnamefont
  {Brower}}, \bibinfo {author} {\bibfnamefont {A.}~\bibnamefont {Hasenfratz}},
  \bibinfo {author} {\bibfnamefont {C.}~\bibnamefont {Rebbi}}, \bibinfo
  {author} {\bibfnamefont {E.}~\bibnamefont {Weinberg}}, \ and\ \bibinfo
  {author} {\bibfnamefont {O.}~\bibnamefont {Witzel}},\ }\href {\doibase
  10.1103/PhysRevD.93.075028} {\bibfield  {journal} {\bibinfo  {journal} {Phys.
  Rev.}\ }\textbf {\bibinfo {volume} {D93}},\ \bibinfo {pages} {075028}
  (\bibinfo {year} {2016})},\ \Eprint {http://arxiv.org/abs/1512.02576}
  {arXiv:1512.02576 [hep-ph]} \BibitemShut {NoStop}%
\bibitem [{\citenamefont {Hasenfratz}\ \emph
  {et~al.}(2017{\natexlab{a}})\citenamefont {Hasenfratz}, \citenamefont
  {Rebbi},\ and\ \citenamefont {Witzel}}]{Hasenfratz:2016gut}%
  \BibitemOpen
  \bibfield  {author} {\bibinfo {author} {\bibfnamefont {A.}~\bibnamefont
  {Hasenfratz}}, \bibinfo {author} {\bibfnamefont {C.}~\bibnamefont {Rebbi}}, \
  and\ \bibinfo {author} {\bibfnamefont {O.}~\bibnamefont {Witzel}},\ }\href
  {\doibase 10.1016/j.physletb.2017.07.058} {\bibfield  {journal} {\bibinfo
  {journal} {Phys. Lett.}\ }\textbf {\bibinfo {volume} {B773}},\ \bibinfo
  {pages} {86} (\bibinfo {year} {2017}{\natexlab{a}})},\ \Eprint
  {http://arxiv.org/abs/1609.01401} {arXiv:1609.01401 [hep-ph]} \BibitemShut
  {NoStop}%
\bibitem [{\citenamefont {Aoki}\ \emph {et~al.}(2012)\citenamefont {Aoki},
  \citenamefont {Aoyama}, \citenamefont {Kurachi}, \citenamefont {Maskawa},
  \citenamefont {Nagai}, \citenamefont {Ohki}, \citenamefont {Shibata},
  \citenamefont {Yamawaki},\ and\ \citenamefont {Yamazaki}}]{Aoki:2012eq}%
  \BibitemOpen
  \bibfield  {author} {\bibinfo {author} {\bibfnamefont {Y.}~\bibnamefont
  {Aoki}}, \bibinfo {author} {\bibfnamefont {T.}~\bibnamefont {Aoyama}},
  \bibinfo {author} {\bibfnamefont {M.}~\bibnamefont {Kurachi}}, \bibinfo
  {author} {\bibfnamefont {T.}~\bibnamefont {Maskawa}}, \bibinfo {author}
  {\bibfnamefont {K.-i.}\ \bibnamefont {Nagai}}, \bibinfo {author}
  {\bibfnamefont {H.}~\bibnamefont {Ohki}}, \bibinfo {author} {\bibfnamefont
  {A.}~\bibnamefont {Shibata}}, \bibinfo {author} {\bibfnamefont
  {K.}~\bibnamefont {Yamawaki}}, \ and\ \bibinfo {author} {\bibfnamefont
  {T.}~\bibnamefont {Yamazaki}} (\bibinfo {collaboration} {{LatKMI}}),\ }\href
  {\doibase 10.1103/PhysRevD.86.059903, 10.1103/PhysRevD.86.054506} {\bibfield
  {journal} {\bibinfo  {journal} {Phys. Rev.}\ }\textbf {\bibinfo {volume}
  {D86}},\ \bibinfo {pages} {054506} (\bibinfo {year} {2012})},\ \Eprint
  {http://arxiv.org/abs/1207.3060} {arXiv:1207.3060 [hep-lat]} \BibitemShut
  {NoStop}%
\bibitem [{\citenamefont {Fodor}\ \emph {et~al.}(2011)\citenamefont {Fodor},
  \citenamefont {Holland}, \citenamefont {Kuti}, \citenamefont {Nogradi},
  \citenamefont {Schroeder}, \citenamefont {Holland}, \citenamefont {Kuti},
  \citenamefont {Nogradi},\ and\ \citenamefont {Schroeder}}]{Fodor:2011tu}%
  \BibitemOpen
  \bibfield  {author} {\bibinfo {author} {\bibfnamefont {Z.}~\bibnamefont
  {Fodor}}, \bibinfo {author} {\bibfnamefont {K.}~\bibnamefont {Holland}},
  \bibinfo {author} {\bibfnamefont {J.}~\bibnamefont {Kuti}}, \bibinfo {author}
  {\bibfnamefont {D.}~\bibnamefont {Nogradi}}, \bibinfo {author} {\bibfnamefont
  {C.}~\bibnamefont {Schroeder}}, \bibinfo {author} {\bibfnamefont
  {K.}~\bibnamefont {Holland}}, \bibinfo {author} {\bibfnamefont
  {J.}~\bibnamefont {Kuti}}, \bibinfo {author} {\bibfnamefont {D.}~\bibnamefont
  {Nogradi}}, \ and\ \bibinfo {author} {\bibfnamefont {C.}~\bibnamefont
  {Schroeder}},\ }\href {\doibase 10.1016/j.physletb.2011.07.037} {\bibfield
  {journal} {\bibinfo  {journal} {Phys. Lett.}\ }\textbf {\bibinfo {volume}
  {B703}},\ \bibinfo {pages} {348} (\bibinfo {year} {2011})},\ \Eprint
  {http://arxiv.org/abs/1104.3124} {arXiv:1104.3124 [hep-lat]} \BibitemShut
  {NoStop}%
\bibitem [{\citenamefont {Cheng}\ \emph {et~al.}(2014)\citenamefont {Cheng},
  \citenamefont {Hasenfratz}, \citenamefont {Liu}, \citenamefont
  {Petropoulos},\ and\ \citenamefont {Schaich}}]{Cheng:2013xha}%
  \BibitemOpen
  \bibfield  {author} {\bibinfo {author} {\bibfnamefont {A.}~\bibnamefont
  {Cheng}}, \bibinfo {author} {\bibfnamefont {A.}~\bibnamefont {Hasenfratz}},
  \bibinfo {author} {\bibfnamefont {Y.}~\bibnamefont {Liu}}, \bibinfo {author}
  {\bibfnamefont {G.}~\bibnamefont {Petropoulos}}, \ and\ \bibinfo {author}
  {\bibfnamefont {D.}~\bibnamefont {Schaich}},\ }\href {\doibase
  10.1103/PhysRevD.90.014509} {\bibfield  {journal} {\bibinfo  {journal}
  {Phys.Rev.}\ }\textbf {\bibinfo {volume} {D90}},\ \bibinfo {pages} {014509}
  (\bibinfo {year} {2014})},\ \Eprint {http://arxiv.org/abs/1401.0195}
  {arXiv:1401.0195 [hep-lat]} \BibitemShut {NoStop}%
\bibitem [{\citenamefont {Aoki}\ \emph {et~al.}(2013)\citenamefont {Aoki},
  \citenamefont {Aoyama}, \citenamefont {Kurachi}, \citenamefont {Maskawa},
  \citenamefont {Nagai}, \citenamefont {Ohki}, \citenamefont {Rinaldi},
  \citenamefont {Shibata}, \citenamefont {Yamawaki},\ and\ \citenamefont
  {Yamazaki}}]{Aoki:2013zsa}%
  \BibitemOpen
  \bibfield  {author} {\bibinfo {author} {\bibfnamefont {Y.}~\bibnamefont
  {Aoki}}, \bibinfo {author} {\bibfnamefont {T.}~\bibnamefont {Aoyama}},
  \bibinfo {author} {\bibfnamefont {M.}~\bibnamefont {Kurachi}}, \bibinfo
  {author} {\bibfnamefont {T.}~\bibnamefont {Maskawa}}, \bibinfo {author}
  {\bibfnamefont {K.-i.}\ \bibnamefont {Nagai}}, \bibinfo {author}
  {\bibfnamefont {H.}~\bibnamefont {Ohki}}, \bibinfo {author} {\bibfnamefont
  {E.}~\bibnamefont {Rinaldi}}, \bibinfo {author} {\bibfnamefont
  {A.}~\bibnamefont {Shibata}}, \bibinfo {author} {\bibfnamefont
  {K.}~\bibnamefont {Yamawaki}}, \ and\ \bibinfo {author} {\bibfnamefont
  {T.}~\bibnamefont {Yamazaki}} (\bibinfo {collaboration} {LatKMI}),\ }\href
  {\doibase 10.1103/PhysRevLett.111.162001} {\bibfield  {journal} {\bibinfo
  {journal} {Phys. Rev. Lett.}\ }\textbf {\bibinfo {volume} {111}},\ \bibinfo
  {pages} {162001} (\bibinfo {year} {2013})},\ \Eprint
  {http://arxiv.org/abs/1305.6006} {arXiv:1305.6006 [hep-lat]} \BibitemShut
  {NoStop}%
\bibitem [{\citenamefont {Olive}\ \emph {et~al.}(2014)\citenamefont {Olive}
  \emph {et~al.}}]{Agashe:2014kda}%
  \BibitemOpen
  \bibfield  {author} {\bibinfo {author} {\bibfnamefont {K.~A.}\ \bibnamefont
  {Olive}} \emph {et~al.} (\bibinfo {collaboration} {Particle Data Group}),\
  }\href {\doibase 10.1088/1674-1137/38/9/090001} {\bibfield  {journal}
  {\bibinfo  {journal} {Chin. Phys.}\ }\textbf {\bibinfo {volume} {C38}},\
  \bibinfo {pages} {090001} (\bibinfo {year} {2014})}\BibitemShut {NoStop}%
\bibitem [{\citenamefont {L{\"{u}}scher}(2010)}]{Luscher:2010iy}%
  \BibitemOpen
  \bibfield  {author} {\bibinfo {author} {\bibfnamefont {M.}~\bibnamefont
  {L{\"{u}}scher}},\ }\href {\doibase 10.1007/JHEP08(2010)071} {\bibfield
  {journal} {\bibinfo  {journal} {JHEP}\ }\textbf {\bibinfo {volume} {1008}},\
  \bibinfo {pages} {071} (\bibinfo {year} {2010})},\ \Eprint
  {http://arxiv.org/abs/1006.4518} {arXiv:1006.4518 [hep-lat]} \BibitemShut
  {NoStop}%
\bibitem [{\citenamefont {Kaneko}\ \emph {et~al.}(2014)\citenamefont {Kaneko},
  \citenamefont {Aoki}, \citenamefont {Cossu}, \citenamefont {Fukaya},
  \citenamefont {Hashimoto},\ and\ \citenamefont {Noaki}}]{Kaneko:2013jla}%
  \BibitemOpen
  \bibfield  {author} {\bibinfo {author} {\bibfnamefont {T.}~\bibnamefont
  {Kaneko}}, \bibinfo {author} {\bibfnamefont {S.}~\bibnamefont {Aoki}},
  \bibinfo {author} {\bibfnamefont {G.}~\bibnamefont {Cossu}}, \bibinfo
  {author} {\bibfnamefont {H.}~\bibnamefont {Fukaya}}, \bibinfo {author}
  {\bibfnamefont {S.}~\bibnamefont {Hashimoto}}, \ and\ \bibinfo {author}
  {\bibfnamefont {J.}~\bibnamefont {Noaki}} (\bibinfo {collaboration}
  {JLQCD}),\ }\href@noop {} {\bibfield  {journal} {\bibinfo  {journal} {PoS}\
  }\textbf {\bibinfo {volume} {LATTICE2013}},\ \bibinfo {pages} {125} (\bibinfo
  {year} {2014})},\ \Eprint {http://arxiv.org/abs/1311.6941} {arXiv:1311.6941
  [hep-lat]} \BibitemShut {NoStop}%
\bibitem [{\citenamefont {Cossu}\ \emph {et~al.}()\citenamefont {Cossu} \emph
  {et~al.}}]{IroIro++}%
  \BibitemOpen
  \bibfield  {author} {\bibinfo {author} {\bibfnamefont {G.}~\bibnamefont
  {Cossu}} \emph {et~al.},\ }\href {https://github.com/coppolachan/IroIro}
  {\enquote {\bibinfo {title} {Iroiro++},}\ }\BibitemShut {NoStop}%
\bibitem [{\citenamefont {Boyle}\ \emph {et~al.}()\citenamefont {Boyle},
  \citenamefont {Cossu}, \citenamefont {Portelli},\ and\ \citenamefont
  {Yamaguchi}}]{GRID}%
  \BibitemOpen
  \bibfield  {author} {\bibinfo {author} {\bibfnamefont {P.}~\bibnamefont
  {Boyle}}, \bibinfo {author} {\bibfnamefont {G.}~\bibnamefont {Cossu}},
  \bibinfo {author} {\bibfnamefont {A.}~\bibnamefont {Portelli}}, \ and\
  \bibinfo {author} {\bibfnamefont {A.}~\bibnamefont {Yamaguchi}},\ }\href
  {https://github.com/paboyle/Grid} {\enquote {\bibinfo {title} {Grid},}\
  }\BibitemShut {NoStop}%
\bibitem [{\citenamefont {Boyle}\ \emph {et~al.}(2015)\citenamefont {Boyle},
  \citenamefont {Yamaguchi}, \citenamefont {Cossu},\ and\ \citenamefont
  {Portelli}}]{Boyle:2015tjk}%
  \BibitemOpen
  \bibfield  {author} {\bibinfo {author} {\bibfnamefont {P.}~\bibnamefont
  {Boyle}}, \bibinfo {author} {\bibfnamefont {A.}~\bibnamefont {Yamaguchi}},
  \bibinfo {author} {\bibfnamefont {G.}~\bibnamefont {Cossu}}, \ and\ \bibinfo
  {author} {\bibfnamefont {A.}~\bibnamefont {Portelli}},\ }\href@noop {}
  {\bibfield  {journal} {\bibinfo  {journal} {PoS}\ }\textbf {\bibinfo {volume}
  {LATTICE2015}},\ \bibinfo {pages} {023} (\bibinfo {year} {2015})},\ \Eprint
  {http://arxiv.org/abs/1512.03487} {arXiv:1512.03487 [hep-lat]} \BibitemShut
  {NoStop}%
\bibitem [{\citenamefont {Hasenfratz}\ \emph
  {et~al.}(2017{\natexlab{b}})\citenamefont {Hasenfratz}, \citenamefont
  {Rebbi},\ and\ \citenamefont {Witzel}}]{Hasenfratz:2017qyr}%
  \BibitemOpen
  \bibfield  {author} {\bibinfo {author} {\bibfnamefont {A.}~\bibnamefont
  {Hasenfratz}}, \bibinfo {author} {\bibfnamefont {C.}~\bibnamefont {Rebbi}}, \
  and\ \bibinfo {author} {\bibfnamefont {O.}~\bibnamefont {Witzel}},\
  }\href@noop {} {\  (\bibinfo {year} {2017}{\natexlab{b}})},\ \Eprint
  {http://arxiv.org/abs/1710.11578} {arXiv:1710.11578 [hep-lat]} \BibitemShut
  {NoStop}%
\bibitem [{\citenamefont {Chiu}(2016)}]{Chiu:2016uui}%
  \BibitemOpen
  \bibfield  {author} {\bibinfo {author} {\bibfnamefont {T.-W.}\ \bibnamefont
  {Chiu}},\ }\href@noop {} {\  (\bibinfo {year} {2016})},\ \Eprint
  {http://arxiv.org/abs/1603.08854} {arXiv:1603.08854 [hep-lat]} \BibitemShut
  {NoStop}%
\bibitem [{\citenamefont {Chiu}(2017)}]{Chiu:2017kza}%
  \BibitemOpen
  \bibfield  {author} {\bibinfo {author} {\bibfnamefont {T.-W.}\ \bibnamefont
  {Chiu}},\ }\href@noop {} {\bibfield  {journal} {\bibinfo  {journal} {PoS}\
  }\textbf {\bibinfo {volume} {LATTICE2016}},\ \bibinfo {pages} {228} (\bibinfo
  {year} {2017})}\BibitemShut {NoStop}%
\bibitem [{\citenamefont {Fodor}\ \emph {et~al.}(2018)\citenamefont {Fodor},
  \citenamefont {Holland}, \citenamefont {Kuti}, \citenamefont {Nogradi},\ and\
  \citenamefont {Wong}}]{Fodor:2017gtj}%
  \BibitemOpen
  \bibfield  {author} {\bibinfo {author} {\bibfnamefont {Z.}~\bibnamefont
  {Fodor}}, \bibinfo {author} {\bibfnamefont {K.}~\bibnamefont {Holland}},
  \bibinfo {author} {\bibfnamefont {J.}~\bibnamefont {Kuti}}, \bibinfo {author}
  {\bibfnamefont {D.}~\bibnamefont {Nogradi}}, \ and\ \bibinfo {author}
  {\bibfnamefont {C.~H.}\ \bibnamefont {Wong}},\ }\href {\doibase
  10.1016/j.physletb.2018.02.008} {\bibfield  {journal} {\bibinfo  {journal}
  {Phys. Lett.}\ }\textbf {\bibinfo {volume} {B779}},\ \bibinfo {pages} {230}
  (\bibinfo {year} {2018})},\ \Eprint {http://arxiv.org/abs/1710.09262}
  {arXiv:1710.09262 [hep-lat]} \BibitemShut {NoStop}%
\end{thebibliography}%
\bibliographystyle{apsrev4-1}

\end{document}